\documentclass[prd,twocolumn,superscriptaddress,showpacs,nofootinbib,
preprintnumbers]{revtex4}

\usepackage{amsmath}
\usepackage{amsfonts}
\usepackage{graphicx}
\usepackage{dcolumn}

\def\be{\begin{equation}}
\def\ee{\end{equation}}
\def\ba{\begin{eqnarray}}
\def\ea{\end{eqnarray}}
\def\bs{\begin{subequations}}
\def\es{\end{subequations}}

\usepackage{color}

\begin{document}

\title{How delicate are the $f(R)$ gravity models with disappearing \\
cosmological constant?}
%______________________________________
\author{I. Thongkool}
\affiliation{Centre for Theoretical Physics, Jamia Millia Islamia,
New Delhi-110025, India}
\author{M. Sami}
\affiliation{Centre for Theoretical Physics, Jamia Millia Islamia,
New Delhi-110025, India}

\author{S. Rai Choudhury}
\affiliation{Centre for Theoretical Physics, Jamia Millia Islamia,
New Delhi-110025, India}

\begin{abstract}
We consider stability of spherically symmetric solutions in $f(R)$
gravity model proposed by Starobinsky. We find that the model
suffers from a severe fine tuning problem when applied to compact
objects like neutron stars. The problem can be remedied by
introducing a cut off on the mass of the scalar degree of freedom
present in the model. A new mass scale associated with neutron stars
density is then required for the stabilities of $f(R)$ gravity
solutions inside relativistic stars.
\end{abstract}

\maketitle

\section{Introduction}
The cause of cosmic repulsion responsible for late time acceleration
is one of the mysteries of modern cosmology. The simplest
possibility to account for this effect is related to the assumption
of dark energy\cite{rev}. It is, however, quite possible that late
time acceleration of universe is the result of large scale
modification of gravity. Amongst all the schemes of modification of
gravity in the infrared regime, the f(R) theories of gravity
\cite{FRB}are most elegant and promising. These theories apart from
the spin-2 object necessarily contain a scalar degree of freedom
dubbed {\it scalaron}. Stability of the theory requires that the
scalaron is not tachyon and graviton is not a ghost which can be
ensured by demanding the positivity of the first and the second
derivatives of $f(R)$ with respect to the Ricci scalar $R$.

The local gravity constraints impose most stringent restrictions on
any scheme of large scale modification of gravity in particular the
$f(R)$ theories of gravity. Most of the $f(R)$ models are either not
cosmologically viable or simply reduce to $\Lambda CDM$\cite{Am}. An
interesting class of models proposed by Hu-Sawicki and Starobinsky
,referred to as HSS hereafter, can reconcile with local gravity
constraints and has the potential capability of being distinguished
from cosmological constant\cite{hufr,star,gannouji}(See
Ref.\cite{Apl1} on the related theme).

The de-Sitter minimum in these models is very near to the curvature
singularity which the scalaron can easily hit while evolving towards
the minimum of its
potential\cite{Apl2,frolov,TH,maeda1,maeda2,NOOD,Lango,whu}. One can
try to modify HSS models by creating a large potential barrier
between the de-Sitter minimum and curvature singularity\cite{waga}.
However, the latter results are in clear violation of local gravity
constraints\cite{Itz}. Thus the presence of finite time singularity
is generic in these models and should be handled carefully. The safe
passage of scalaron to the minimum of its potential requires fine
tuning of its initial conditions. The situation gets worse in the
high curvature regime.

In this paper we examine the fine tuning problem associated with
$f(R)$ theories with disappearing cosmological constant. We
demonstrate that the level of fine tuning the HSS models require in
case of the compact objects like neutron stars poses serious
challenge to these models. The problem can be alleviated by
introducing the quadratic curvature terms in the HSS scenario.

\section{ The scalar degree of freedom in $f(R)$ theories of gravity}
We consider the modification of the Einstein-Hilbert action in the
presence of the matter Lagrangian $\mathcal{L}_m$
\begin{equation}
 \mathcal{S} = \int d^4 x \sqrt{-g} \left [ \frac{f(R)}{2} + \mathcal{L}_m
\right ]
\end{equation}
where $f(R)$ is a function of the Ricci scalar $R$. Variation with respect to
metric leads to the following field equations
\begin{equation}
\label{freqn}
 f_{,R} R_{\mu\nu} - \nabla_{\mu}\nabla_{\nu} f_{,R}
+\left ( \Box f_{,R} - \frac{1}{2} f \right ) g_{\mu\nu}
= T_{\mu\nu},
\end{equation}
where $f_{,R} \equiv df/dR$ and
$T_{\mu\nu} \equiv -2\delta \mathcal{L}_m/\delta g^{\mu\nu}
+ g_{\mu\nu} \mathcal{L}_m$.
 The $f(R)$ theories of gravity necessarily contain an additional scalar degree of freedom
  which becomes clear after taking the trace of $(\ref{freqn})$
\begin{equation}
 \Box \psi = \frac{1}{3} T + \frac{dV}{d\psi},
 \label{eq:trace}
\end{equation}
where $\psi \equiv f_{,R} ;~~   dV/d\psi = \left ( 2 f(R(
\psi))-\psi R(\psi)\right)/3$.

In what follows, it would be convenient to us to write $f(R)$ in the
form of the correction term to Einstein-Hilbert action, $\Delta$,
\begin{equation}
 f(R)=R+\Delta, \qquad \psi = 1 + \Delta_{,R},
\end{equation}
where $\Delta_{,R}$ denotes the derivative of the correction term
with respect to the Ricci scalar $R$.

In case of Starobinsky  model \cite{star}, we have
\begin{eqnarray}
&& \Delta = -\lambda R_c \left [ 1- \left (1+R^2/R_c^2 \right )^{-n}
\right ]\\
&&\psi = 1-2n\lambda \left ( \frac{R_c}{R} \right )^{2n+1} \quad
\textrm{for} \quad R \gg R_c.
\end{eqnarray}

Taking $R_c = \rho_{_{\Lambda}}/4$ where $\rho_{_{\Lambda}}$ is the
cosmological density, $(10^{-29}$ g/cm$^3)$ and $R \sim \rho_c$ is
the curvature inside the neutron star such that $(\rho_c \sim
10^{14}$g/cm$^3)$. The numerical value of $\psi_0$ corresponding to
the de-Sitter minimum is given by
\begin{equation}
 \psi_0 \approx \left \{
\begin{array}{ll}
 1 - \mathcal{O}\left (10^{-122} \right ) \quad n=0.9,\\
\\
 1 - \mathcal{O}\left (10^{-217} \right ) \quad n=2.
\end{array} \right.
\end{equation}

The Local Gravity Constraints  are satisfied for $n \gtrsim 0.9$
\cite{shinjichm} and the evolution of density perturbations during
the matter-dominated epoch requires $n \gtrsim 2$ for
$G_{eff}=G/f'(R)$ to be consistent with observations \cite{star}.

The scalar degree of freedom plays an important role in $f(R)$
theories of gravity, namely, its dynamics controls the space-time
curvature. In generic cases, the de Sitter minimum at $\psi_0$  is
very close to $\psi = 1$  corresponding to curvature singularity.
The finite barrier between the singularity and de Sitter minimum
means the curvature singularity is energetically accessible. In
order to avoid it, the evolution of the field needs to be fine
tuned. We shall see later that in the case of the neutron star with
constant density $\rho_c$, the extreme fine tuning of initial
conditions becomes necessary for the existence of GR-like solution
$(R \sim \rho_c)$ along the radius $r$ of the star to match the
correct boundary conditions at its surface, $r=r_{\ast}$.

\section{The growing mode of the perturbation and fine tuning of initial conditions}
The problem of fine tuning is inherent in $f(R)$ theories if they
are to be consistent with local gravity constraints and can be
appreciated by using the analytical arguments. An approximation
scheme for the solution of $\psi(r)$ can be set up using the
iterative computation for $R(r)$ as follows\cite{hufr,star},
\begin{equation}
 R(r) = R_0(r) + \delta R_1(r) + \delta R_2(r)+\cdots,
\end{equation}
where $R_0(r) = \rho(r)-3P(r)$ and the first order iteration gives
rise to the following expression,
\begin{equation}
 \delta R_1 = \left [ -3 \nabla^2 \Delta_{,R} + \Delta_{,R}R-2\Delta \right ] \big |_{R=R_0}.
\end{equation}
In case of Starobinsky model, we have
\begin{eqnarray}
 &&|\nabla^2 \Delta_{,R}|,|\Delta_{,R}R| \ll |\Delta| \qquad
\textrm{for} \quad R \gg R_c,\\
&& \delta R_1(r) \approx -2\Delta \approx 2 \lambda R_c = const.
\end{eqnarray}
In the first order iteration, $\psi_1$ can be expressed through the
GR-like solution $\psi_0$ as
\begin{eqnarray}
 \psi_1 &=& 1+\Delta_{,R}|_{R=R_0+\delta R_1},\\
&\approx& 1 - 2n\lambda \left ( \frac{R_c}{R_0} \right )^{2n+1}
\left [ 1 + \delta R_1/R_0 \right ]^{-2n-1}, \\
&\approx& \psi_0+4n\lambda^2(2n+1)\left ( \frac{R_c}{R_0} \right )^{2n+2},
\end{eqnarray}
where we have used $\delta R_1  = 2 \lambda R_c$.

Let us note that the scalaron mass in high curvature regime $R
\approx R_0 \approx \rho \gg R_c$ is given by
\begin{equation}\label{eq:mass}
 m_{\psi_0}^2 = \frac{d^2V}{d\psi^2}|_{\psi=\psi_0}
\approx \frac{R_c}{6n(2n+1)\lambda} \left ( \frac{R_0}{R_c} \right )^{2n+2},
\end{equation}
which allows us to obtain the first order iteration solution,
\begin{equation}
 \psi_1(r) = \psi_0(r) + \frac{2}{3}\lambda \frac{R_c}{m_{\psi_0}^2}.
\end{equation}
The GR-like solution, $\psi_1(r)$ under consideration, can deviate
from $\psi_0(r)$ only near the stellar radius otherwise many known
observational constraints of neutron stars will not be satisfied.
Indeed, the first order iteration $\psi_1$ solution is approximately
the Schwarzschild de Sitter solution because it is corresponding to
the curvature $R-2\lambda R_c$ which differs  from $R$ of the GR
solution by a constant. In large scalaron mass limit, $\psi$ reduces
to GR solution as expected.

The configuration of the perturbative solutions of the $f(R)$
gravity and general relativity are very different in the limit of
the large scalaron mass. As demonstrated Refs.\cite{dolgov}
$\&$\cite{hustability}, the sign and the size of $m_{\psi}^2$ play a
crucial role for the stability of solutions in time. The
Dolgov-Kawasaki instability can be avoided by choosing $m_{\psi}^2 >
0$ which makes the perturbation $\delta \psi(t)$ oscillating in
time. However, a large positive $m_{\psi}^2$ causes the instability
of static spherically symmetric solutions \cite{kimmo} for a class
of $f(R)$ gravity models which are carefully built to evade the
local gravity constraints \cite{hufr}\footnote{ This type of
instability that we shall focus on in the subsequent discussion
refers to the radial evolution.}.

To demonstrate the catastrophic instability from a large positive
value $m_{\psi}^2$, let us consider a small perturbation $\delta
\psi_1(r)$ around $\psi_1(r)$. Assuming the static spherically
symmetric metric, the trace equation (\ref{eq:trace}) tells us that
\begin{eqnarray}
 \delta \psi_1''+\frac{2}{r} \delta \psi_1'
&=& \frac{dV}{d\psi} \Big |_{\psi=\psi_1+\delta \psi_1}
-\frac{dV}{d\psi} \Big |_{\psi=\psi_1}, \\
&\approx&\frac{d^2V}{d\psi_1^2}\delta \psi_1 = m_{\psi_1}^2 \delta \psi_1,
\end{eqnarray}
where primes denote derivatives with respect to $r$.

In case $m_{\psi_1}^2 > 0$, the growth of the perturbation $\delta
\psi_1(r)$ along the radius can be obtained in form
\begin{equation}
 \delta \psi_1(r) = \delta\tilde{\psi}_1(0)\left \lbrace C_1 \frac{e^{m_0r}}{r}
+C_2 \frac{e^{-m_0r}}{r} \right \rbrace, \label{PERT}
\end{equation}
where we have used the notation $m_0 \equiv m_{\psi_1} \approx
m_{\psi_0}$. It follows from Eq.(\ref{PERT}) that the exponentially
growing mode of $\delta \psi_1$ is unavoidable. Thus the instability
of solutions always persists for any metric $f(R)$ gravity models
with the large $m_{\psi}$. This fact may be exhibited by casting the
equation of perturbation $\delta \psi_1(t,r)$ in the following form,
\begin{equation}
 \left ( \partial_t^2 - \vec{\nabla}^2 \right ) \delta \psi = -m_0^2 \delta \psi.
\end{equation}
It should be noticed that the difference of the sign of
$\partial_t^2$ and $\vec{\nabla}^2$ comes from the signature of the
metric itself. The avoidance of Dolgov-Kawasaki time instability by
choosing $m_0^2>0$ invokes the instability ($\delta \psi_1(r)
\propto e^{\pm m_0 r}$) of the static solution. The orthogonality of
the stability conditions of the time and space arising from the
signature of the metric was proposed in Ref.\cite{kimmo} and the
evidence of this instability of the static solutions was observed
numerically as the problem of the existence of relativistic stars in
$f(R)$ gravity theories \cite{maeda1},\cite{maeda2}.

It is however difficult but necessary to maintain the small
deviation from the GR-like solution $\psi_0(r)$ for the whole range
of the stellar radius when the growth of perturbation is
exponential. The initial value $\delta \psi_1(0)$ must be extremely
fine tuned if we want to stay near the GR-like solution. The
seriousness of the fine tuning is related to the size of the number
$m_0 r_\ast$ which is typically huge. The length scale corresponding
to $m_0$, the Compton wavelength $\lambda_c \equiv 1/m_0$, is very
small in the nuclear matter density regime due to chameleon effect.
Without the cut off on $m_0$, $\lambda_c$ can shrink below {\it
Planck length}( $r_{_P} = \sqrt{\hbar G /c^3} \sim 10^{-33}$ cm).
Meanwhile, the stellar radius $r_\ast$ is large ($r_\ast \sim 10^6$
cm. for neutron stars) which means that $m_0r_\ast =
r_\ast/\lambda_c \gg 1$.

Let us estimate the number of $m_0 r_\ast$ for a neutron star using
the following relation,
\begin{equation}
 r_{\ast}^2 = \frac{12P_c}{\rho_c^2}=\frac{12\omega}{\rho_c}
\end{equation}
where $P_c$ is the pressure at the center of the neutron star and
the equation of state $P_c = \omega \rho_c$ is assumed.

Using the approximation $R_0 \sim \rho_c$ and  the constant equation
of state parameter, $\omega \sim 0.1$, we can rewrite $r_\ast$ in
the term of Hubble length $r_{_{H_0}} = c/H_0 \sim R_c^{-1/2}$ as
\begin{equation}
r_{\ast} \sim R_0^{-1/2} = \left ( \frac{R_c}{R_0} \right )^{1/2} r_{_{H_0}}
\sim \left (\frac{\rho_{_\Lambda}}{\rho_c} \right )^{1/2} r_{_{H_0}},
\end{equation}
The faster shrinking of $\lambda_c$ via Chameleon effect can be seen
using Eq.(\ref{eq:mass})
\begin{equation}
 \lambda_c = 1/m_0 \sim \left (\frac{\rho_{_\Lambda}}{\rho_c} \right )^{n+1} r_{_{H_0}}.
\end{equation}
which means that the size of $m_0r_\ast$ depends on the density
contrast in the following way
\begin{equation}
 m_0r_\ast = r_\ast/\lambda_c \sim \left (\frac{\rho_c}{\rho_{_\Lambda}} \right )^{n+1/2}
\sim 10^{43(n+1/2)}.
\end{equation}
With the minimum requirement $n \ge 0.9$ for local gravity
constraints, the growing mode at $r=r_{\ast}$ becomes
\begin{equation}
\delta \psi_1(r_\ast) = \frac{\delta \tilde{\psi}_1(0)C_1}{r_\ast} \exp[10^{60}].
\end{equation}
If $C_1 \ne 0$ or the growing mode is allowed, the initial condition
$\delta \tilde{\psi}_1(0)$ must be fine tuned to a fantastic level
in order to compensate the enormous factor $\exp[10^{60}]$!

As we have shown, this tuning problem arises from the same criterion
as Dolgov-Kawasaki time instability. The allowance of the incredibly
small set of the initial condition corresponding to the correct
boundary condition of GR-like solution should be considered as a
serious theoretical problem. The system which is highly sensitive to
the initial value to the level of
$\mathcal{O}(10^{60}\exp[10^{-60}])$ should not be considered as
satisfactory.

Let us note that setting $C_1 = 0$ from the beginning is not
permissible because the continuity of the gradient of the solution
$\psi(r) = \psi_1(r) + \delta \psi_1(r)$ at the center of the star
needs
\begin{equation}
 \psi_1'(0) + \delta \psi_1'(0) = \psi_0'(0)+ \delta \psi_1'(0) = 0,
\end{equation}
which leads to $\delta \psi_1'(r) = 0$ or $C_1 = -C_2$ whereas the
for GR-like solution to hold, $\psi_1'(0)=\psi_0'(0) = 0$ and $m_0
\approx const$ are assumed. Then the perturbation $\delta \psi_1(r)$
can be rewritten as
\begin{equation}
 \delta \psi_1 (r) = \delta \psi_1(0) \frac{\sinh(m_0 r)}{m_0 r},
\end{equation}
which gives the evolution of perturbation from centre to the surface
of the star. We can estimate how the initial perturbation,$\delta
\psi_1(0)$, at the center enhances as we move to the surface of the
star,
\begin{equation}
\delta \psi_1 (r_\ast) \approx \delta \psi_1(0)
\frac{\exp[10^{60}]}{2\times10^{60}},
\end{equation}
The situation becomes worse for the growth of perturbations in case
$n \gtrsim 2$ with $m_0 r_\ast \sim 10^{107}$.
%\subsection{The example of the tuning initial condition}

There are two ways to deviate from the first order iteration
solution $\psi_1(0)$ depending on the initial sign of $\delta
\psi_1(0)$. For the positive $\delta \psi_1(0)$, the perturbation
drives $\psi (r)$ toward curvature singularity $\psi(r) = 1$ while
in the case of $\delta \psi_1(0) < 0$, one can easily  get
 $\psi(r)$ which is inconsistent with the observational constraints. Since $\psi(r)$
 and $R(r)$ are in one to one correspondence by definition, the GR-like solution $\psi(r)$
 cannot much depart from the value $\psi_0(r)$ and $\psi_1(r)$ for
the entire stellar radius otherwise many known constraints of
general relativity  such as the one coming from double pulsar tests
would not be satisfied.

The simplest way to satisfy stringent test of general
 relativity is provided by taking the  Schwarzschild
 de-Sitter solution with small cosmological constant, $2\lambda R_c$.
 From the $1^{st}$ iteration, the small interval of allowed deviation  is
\begin{eqnarray}
 \psi_1(r)-\psi_0(r) &\approx& 4n\lambda^2(2n+1)\left ( \frac{R_c}{R_0} \right )^{2n+2}, \\
&\sim& \left ( \frac{\rho_{_{\Lambda}}}{\rho_c} \right )^{2n+2},
\end{eqnarray}
where we have assumed $n,\lambda \sim \mathcal{O}(1)$.

In case $\delta \psi_1(0)<0$, the maximum allowed deviation at
$r_\ast$ can be approximated by
\begin{equation}
 \delta \psi_1(0)\frac{\exp[m_0r_\ast]}{2m_0r_\ast}
\gtrsim -\left ( \frac{\rho_{_{\Lambda}}}{\rho_c} \right )^{2n+2}.
\end{equation}
This requirement for GR-like solutions hold leads to an extreme fine
tuning, for example, in case $n=0.9$ ($ m_0r_\ast=10^{60}$), we need
$|\delta\psi_1(0)| \lesssim 6.3\times10^{-104}\exp[-10^{60}]$. With
this huge size of the scalaron mass inside neutron stars, the
GR-like solution is highly unstable. Any small perturbation from the
exact solution, no matter how small in the physical sense, can cause
a catastrophic effect such as the divergence of $R$ or the solution
cannot be compatible with observations\cite{maeda1}.

\section{ Scalaron mass cut off induced by $(\mu/R_c) R^2$ term}
The fine tuning problem in $f(R)$ theories is closely associated
with the large mass, the scalaron acquires in the high curvature
regime which is essential for local gravity constraints to be
evaded. It is, however, unsatisfactory that the scalaron mass can
easily exceed the Planck mass\cite{star}. Using the dimensional
arguments, we can estimate the stellar radius of neutron star from
the nuclear matter density, cosmological density and the Hubble
radius,
\begin{equation}\label{eq:lengthscale}
 \frac{r_\ast}{r_{_{H}}} = \left ( \frac{\rho_{_\Lambda}}{\rho_c} \right )^{1/2}
\quad \textrm{or} \quad r_\ast = 10^{-43/2}\times1.3\times10^{26}
\textrm{m}.
\end{equation}
which gives the correct order of magnitude for the radius of neutron
star, $\mathcal{O}(10^4)$ m.  It is clear from Eq.(\ref{eq:mass})
that mass scale (length scale) corresponding to the scalaron mass
increases (decreases) faster with density  for any $n>0$,
\begin{equation}\label{eq:mass2}
 m_{\psi} \sim \sqrt{R_c} \left ( \frac{\rho}{\rho_{_\Lambda}} \right )^{n+1}.
\end{equation}
As pointed out in Ref.\cite{star},  the scalaron mass can exceed the
Planck mass even in the regime of the density of classical
relativity. Let us estimate $\sqrt{R_c}$ in the unit of $M_p$,
\begin{equation}\label{eq:mp}
 \frac{\sqrt{R_c}}{M_p} \sim \frac{H_0}{M_p}
= \frac{2.13 h\times 10^{-42} GeV/\hbar}
{1.22\times 10^{19} GeV/c^2}
\sim 10^{-61}.
\end{equation}
From Eq.(\ref{eq:mass2}) and Eq. (\ref{eq:mp}), we  find that
$m_\psi \geq M_p$ when $\rho \gtrsim 10^{32} \rho_{_\Lambda}$ for $n
= 0.9$ and $\rho \gtrsim 10^{20} \rho_{_\Lambda}$ for $n = 2$.
Around the Big Bang nucleosynthesis (BBN), the density
$\rho_{_{BBN}} \sim 10^{30} \rho_{_\Lambda}$ \cite{zhang} giving
rise to $m_\psi \sim 10^{-4}Mp$ for $n=0.9$ which is heavier than
the typical inflaton mass $10^{-6}Mp$. The scalaron mass  exceeds
the Planck mass by the factor $10^{29}$ in case  $n=2$.

In stars with nuclear matter density, $\rho_c \sim 10^{43}
\rho_{_\Lambda}$, $m_{\psi}$ becomes $10^{20.7}M_p$ for $n=0.9$ and
$10^{68}M_p$ for $n=2$. Hence, the length scales $\lambda_c$
corresponding to this un-physical heavy masses are definitely
shorter than Planck length , equivalently, the gigantic numbers,
$m_0 r_\ast = r_\ast / \lambda_c$ is also un-physical due to the
uncontrollable Chameleon effect.

The simplest way to remedy the fine tuning problem is provided by
putting a cut off on $m_{\psi}$ by carefully chosen the maximum
value of $m_\psi$ such $m_0 r_\ast \sim \mathcal{O}(1)$ which can be
achieved by adding $(\mu/R_c)R^2$ term into the model under
consideration\cite{star,TH}.

Indeed, in the limit $(R\gg R_c)$ in context with the Starobinsky
model, we can use the approximation
\begin{eqnarray}
 m_{\psi}^2 &=&\frac{1}{3} \left [ \frac{1}{\Delta_{,RR}}
+ \frac{\Delta_{,R}}{\Delta_{,RR}} -R \right ] \\
&\approx&
\frac{1}{3}(\Delta_{,RR})^{-1}
\end{eqnarray}
where $\Delta_{,RR} = d \Delta_{,R} / dR$. The additional term
$(\mu/R_c)R^2$ can provide a cut off on $m_\psi \sim
(R_c/(6\mu))^{1/2}$ when $\mu$ is chosen to satisfy the
condition\cite{star}
\begin{equation}\label{eq:approxmu}
 \left ( \frac{R_c}{R} \right ) \gg \mu \gg \left ( \frac{R_c}{R} \right )^{2(n+1)}.
\end{equation}
In case of  neutron stars, the upper limit $\mu \ll
\rho_{_\Lambda}/\rho_c \sim 10^{-43}$ is equivalent to $(\mu / R_c)
R^2 \ll R$ or the correction term should be small compared to the
background curvature $R$. It should be noted that the lower limit is
always satisfied for generic values of $n$.

For the numerical calculation, we need not to stick to approximation
(\ref{eq:approxmu}) as $m_\psi$ can be calculated directly. As
demonstrated in Ref.\cite{maeda2}, the carefully chosen $\mu$
corresponding to the cut off on mass about the scale of neutron star
density, can give rise to GR-like solution in for neutron stars. We
may explain this observation by consider the heuristic argument
(\ref{eq:lengthscale}) which gives $r_\ast \sim 10^{-43/2}R_c^{-1/2}
$ and the corresponding Compton wavelength for the  mass cut off,
$\lambda_c \sim 1/m_\psi \sim (6\mu)^{1/2} R_c^{-1/2} $ such that
\begin{equation}
r_\ast \sim \frac{1}{m_0} = \lambda_c \quad \textrm{when} \quad 6\mu\sim 10^{-43}.
\end{equation}
which clearly allows to avoid the catastrophic fine tuning problem
in compact objects like neutron stars.

\section{Conclusions}
We have examined the scalaron dynamics in the frame work of
Starobinsky model. The curvature singularity is generic to this
class of models if they are to be consistent with local gravity
constraints. The finite potential barrier between the de-Sitter
minimum and the curvature singularity is a serious threat to models
with disappearing cosmological constant. The problem becomes grave
in high curvature regime in compact objects like neutron stars. In
this case, the scalaron mass becomes larger than the Planck mass due
to chameleon mechanism necessary for local gravity constraints to be
evaded. This in turn becomes the root cause of instability problem
in the scenario under consideration. While evolving the scalaron
from the centre to the surface along the radius of neutron star, we
need to stay close GR. Little perturbation of initial conditions at
the centre can easily destroy the desired evolution. This condition
for having GR-like solutions requires  extreme fine tuning of
initial conditions, for instance, in case of  $n=0.9,
m_0r_\ast=10^{60}$, we need $|\delta\psi_1(0)| \lesssim
6.3\times10^{-107}\exp[-10^{60}]$. Situation further worsens for
larger values of $n$. This intractable level of fine tuning throws a
serious challenge to $f(R)$ theories consistent with local gravity
tests.

The problem can be alleviated by introducing quadratic curvature
term $\mu R^2/R_0$ in the Starobinsky model which is equivalent to
putting a cut off on scalaron mass corresponding to
 $\mu
\ll \rho_{_\Lambda}/\rho_c \sim 10^{-43}$ for nuclear matter
density. This prescription, however, runs into problem if one asks
for its compatibility with early universe physics {\it a la}
inflation which leads to much larger value of the scalaron mass.

\section{Acknowledgements}
We thank T. Kobayashi for discussion. IT is supported by ICCR and MS
is supported by ICTP.


\begin{thebibliography}{99}
\bibitem{rev}
\mbox{E.~J.~Copeland, M.~Sami and S.~Tsujikawa, Int. J. Mod.}
\mbox{Phys. D{\bf15},1753 (2006) [arXiv:hep-th/0603057]; M. Sami,}
\mbox{arXiv:0904.3445; V.~Sahni and A.~A.~Starobinsky, Int.\ J.}
\ Mod. \ Phys.\ D\textbf{9}, 373 (2000) [arXiv:astro-ph/9904398];
\mbox{T.~Padmanabhan, Phys.\ Rept.~\textbf{380}, 235 (2003) [arXiv:}
\mbox{hep-th/0212290]; E.~V.~Linder, Gen. Rel. Grav. {\bf40}, 329}
\mbox{(2008) [arXiv:astro-ph/0704.2064]; J. Frieman, M. Turner}
\mbox{and D. Huterer, Ann. Rev. Astron. Astrophys.~{\bf46}, 385}
(2008) [arXiv:0803.0982];R. R. Caldwell and M. Kamionkowski,
arXiv:0903.0866; A. Silvestri and M. Trodden, arXiv:0904.0024;
V. Sahni and A.~A.~Starobinsky, Int. J. Mod. Phys. D{\bf15}, 2105 (2006) [arXiv:astro-ph/0610026]; T. Padmanabhan, AIP Conf. Proc.~{\bf 861},
179 (2006) [arXiv:astro-ph/0603114];
\bibitem{FRB}
S.~Capozziello, Int.\ J.\ Mod.\ Phys.\  D{\bf11}, 483 (2002) [arXiv:gr-qc/0201033];
S.~Capozziello, S.~Carloni, and A.~Troisi
Recent. Res. Dev. Astron. Astrophys.~{\bf1}, 625 (2003) [arXiv:astro-ph/0303041];
S.~Carroll, V.~Duvvuri, M.~Trodden and M.~S.~Turner, Phys.\ Rev.\
D{\bf70}, 043528 (2004) [arXiv:astro-ph/0306438];
T.~P.~Sotiriou and  V.~Faraoni [arXiv:0805.1726];
S.~Nojiri and S.~Odintsov, Gen.\ Rel.\ Grav.\ {\bf 36}, 1765 (2004)
[arXiv:hep-th/0308176].
\bibitem{Am}
L. Amendola, R. Gannouji, D. Polarski and S.
Tsujikawa, Phys. Rev.D{\bf75}, 083504 (2007) [arXiv:gr-qc/0612180].
\bibitem{hufr}
Wayne ~Hu and I.~Sawicki, Phys.\ Rev.\ D{\bf76}, 064004 (2007)
[arXiv:0705.1158].
\bibitem{star}
A. A. Starobinsky, JETP. Lett. {\bf 86}, 157 (2007), [arXiv:0706.2041].
\bibitem{gannouji}
R. Gannouji, B. Moraes and D. Polarski
{\bf JCAP 0902}, 034 (2009) [arXiv:0809.3374].
%%%%%
\bibitem{Apl1}
S. A. Appleby and R. A. Battye,
Phys. Lett. B{\bf654}, 7 (2007) [arXiv:0705.3199].
\bibitem{Apl2}
S. A. Appleby and R. A. Battye, {\bf JCAP 0805}, 019 (2008) [arXiv:0803.1081].
\bibitem{frolov}
A. V. Frolov, Phys. Rev. Lett.~{\bf101}, 061103 (2008) [arXiv:0803.2500].
\bibitem{TH}
Abha Dev , D. Jain  , S. Jhingan  , S. Nojiri , M. Sami and
I. Thongkool, Phys. Rev. D{\bf 78}, 083515 (2008) [arXiv:0807.3445].
\bibitem{maeda1}
 T. Kobayashi and K. Maeda, Phys. Rev. D{\bf78}, 064019 (2008) [arXiv:0807.2503].
\bibitem{maeda2}
T. Kobayashi and K. Maeda, Phys. Rev. D{\bf79}, 024009 (2009) [arXiv:0810.5664].
\bibitem{NOOD} S. Nojiri and S. D. Odintsov, arXiv:0903.5231; S. Nojiri and S. D.
Odintsov, Phys. Rev. D{\bf78}, 046006 (2008) [arXiv:0804.3519];
K. Bamba, S. Nojiri and S. D. Odintsov, {\bf JCAP 0810}, 045 (2008) [arXiv:0807.2575];
\mbox{S. Capozziello, M. De Laurentis , S. Nojiri and S. D.}
\mbox{Odintsov, Phys. Rev. D{\bf79}, 124007 (2009) [arXiv:0903.2753].}
\bibitem{Lango} E. Babichev and D. Langlois, arXiv:0904.1382.
\bibitem{whu} A. Upadhye and Wayne Hu, arXiv:0905.4055.
\bibitem{waga} V. Miranda, S. E. Joras, I. Waga and M. Quartin,
Phys. Rev. Lett. {\bf102}, 221101 (2009) [arXiv:0905.1941].
\bibitem{Itz}
I. Thongkool, M. Sami, R. Gannouji and S. Jhingan, arXiv:0906.2460
\bibitem{shinjichm}
\mbox{S. Capozziello and S. Tsujikawa, Phys.~Rev.D{\bf77},~107501}
(2008) [arXiv:0712.2268].
\bibitem{dolgov}
A.~D.~Dolgov and M. Kawasaki, Phys. Lett. B{\bf573}, 1 (2003)
[arXiv:astro-ph/0307285] .
\bibitem{hustability}
I.~Sawicki~and~Wayne Hu,~Phys.~Rev.~D{\bf 75},~127502~(2007)
[arXiv:astro-ph/0702278.]
\bibitem{kimmo}
K. Kainulainen and D. Sunhede, Phys.~Rev.~D{\bf78},~063511 (2008) [arXiv:0803.0867].
\bibitem{zhang} P. Zhang, Phys. Rev. D{\bf76}, 024007 (2007) [arXiv:astro-ph/0701662].
\end{thebibliography}
\end{document}